\documentstyle[amssymb,preprint,aps]{revtex}
\tightenlines

\begin{document}
\title{Normal State Ettingshausen Effect in La$_{2-x}$Sr$_{x}$CuO$_{4}$}
\author{T.Plackowski and M.Matusiak}
\address{Institute of Low Temperature and Structure Research, \\
Polish Academy of Sciences,\\
50-950 Wroc\l aw, P.O.Box 1410\\
POLAND}
\date{Submitted to Superconductor Science and Technology, April 2, 1999}
\maketitle

\begin{abstract}
A method for measurements of small (metallic) Ettingshausen\ coefficient ($P$%
) was developed. The influence of the dominating thermal effects, the Joule
and Thomson heats, was eliminated making use of the odd symmetry of the
Ettingshausen temperature gradient in respect with reversing of the
direction of the magnetic field and electrical current. The method was
applied to La$_{2-x}$Sr$_{x}$CuO$_{4}$ ($x=0.03\div 0.35$) high-T$_{c}$
superconductor in normal state. We have found that in the whole composition
range the Ettingshausen coefficient is of the order of 10$^{-7}$ m$^{3}$K/J
which is characteristic of typical metals.\ The coefficient$\,$changes sign
from positive to negative near $x\approx $ 0.07. Weak variation of $P$ is in
contrast to the behavior of other transport coefficients for La$_{2-x}$Sr$%
_{x}$CuO$_{4}$, as the thermoeletric power or the Hall coefficient, which
have been reported in literature to change their values by more than two
orders of magnitude with Sr doping.
\end{abstract}

\section{Introduction}

Since the discovery of high-$T_{c}$ (HTC) superconductors their normal state
properties are a subject of extensive investigations. Many of them, as the
electrical resistivity $\rho $, the Hall effect $R_{H}$ or the
thermoelectric power $S$, exhibit some universal behavior in function of
charge carrier doping. The universal behavior of the latter quantity is well
known, for many of the HTC families the thermopower changes sign at optimal
carrier concentration \cite{Obertelli}. The linearity of resistivity for the
optimally doped samples in wide temperature region is also one of their most
well known features \cite{Martin}. The Hall coefficient for optimally doped
samples varies as $1/T$, which results in the $1/T^{2}$ dependence of the
Hall mobility, $\mu _{H}$. Then it appeared that the $1/T^{2}$ dependence of 
$\mu _{H}$ is more universal. It applies not only to the optimally doped,
but also to the under- and overdoped materials \cite{Kubo1}.

In this work we present the measurements of the Ettingshausen coefficient,
one of the less known transport coefficients. Our aim was to complement the
present knowledge of the normal state properties of HTC materials, and,
hopefully, to search for some new universalities. The Ettingshausen effect
is a thermal analog of the Hall effect (the definition and sign convention
are shown in Fig. 1). The difference with respect to the Hall effect lies in
the fact that the temperature difference is measured in the direction
perpendicular to both the current and field directions, instead of the
voltage difference.

The Hall and Ettingshausen effects are two of the four transversal
magneto-thermal effects: 
\begin{eqnarray*}
\nabla \Phi _{\bot } &=&R_{H}~\vec{j}\times \vec{B}\text{, the Hall effect,}
\\
\nabla T_{\bot } &=&P~\vec{j}\times \vec{B}\text{, the Ettingshausen effect,}
\\
\nabla \Phi _{\bot } &=&Q~\nabla T\times \vec{B}\text{, the Nernst effect,}
\\
\nabla T_{\bot } &=&S_{RL}~\nabla T\times \vec{B}\text{, the Righi-Leduc
effect,}
\end{eqnarray*}
where $R_{H}$, $P$, $Q$, $S_{RL}$ are the respective coefficients, $\nabla
\Phi _{\bot }$ and $\nabla T_{\bot }$ are the transversal gradients of
electrical potential and temperature, respectively, caused by the presence
of magnetic field ($\vec{B}$) perpendicular to the electrical current ($\vec{%
j}$) or heat flux ($\vec{q}\sim \nabla T$). All these coefficients are
interconnected by three fundamental relations which were considered by
P.W.Bridgman \cite{Bridgman} in terms of thermodynamics of reversible
processes:

\begin{eqnarray*}
Q &=&\frac{\kappa }{T}P\text{,} \\
Q &=&\frac{\mu _{T}}{\rho }R_{H}\text{,} \\
P &=&\frac{\mu _{T}T}{\kappa }S_{RL}\text{,}
\end{eqnarray*}
where $\kappa $ is the total thermal conductivity and $\mu _{T}$ is the
Thomson coefficient .

The theory of the Ettingshausen effect for semiconductors was considered in 
\cite{Paranjape}, and for metals in \cite{Fieber}. Only few measurements of
the Ettingshausen effect in different materials were carried out up to date.
Typical measured values of the Ettingshausen coefficients for semiconductors
are positive and of the order of 10$^{-2}\div $10$^{-4}$ m$^{3}$K/J (Ge:\cite
{Mette1}; Si:\cite{Mette2}; PbSe, PbTe:\cite{Putley}). Typical values of the
Ettingshausen coefficient for metals are much lower then for semiconductors
and are of the order 10$^{-7}\div $10$^{-8}$ m$^{3}$K/J. A negative
Ettingshausen coefficient was observed for Ag, Cd, Cu, Fe, Zn and Au,
whereas positive for Al, Co and Ni. The Ettingshausen effect and the Hall
effect have opposite signs for Al, Cd, Fe, Ni and Zn, whereas the same signs
for Ag, Co, Cu and Au \cite{CriticalTables,Bridgman}. The Ettingshausen
coefficient is much higher for semi-metals ($\sim $10$^{-4}$ m$^{3}$K/J for
Sb and $\sim $10$^{-3}$ m$^{3}$K/J for Bi \cite{Bridgman}) and for rare
earths ($\sim $10$^{-3}$ m$^{3}$K/J for Y, Gd, Tb and Dy \cite{Zecchina}).

According to our knowledge, no measurements of the Ettingshausen effect were
carried out in HTC superconductors in normal state. Few works were devoted
to the normal state Nernst effect: for Tl-2212 \cite{Clayhold}, YCBO \cite
{Gasumyants} and NdCeCuO \cite{Fournier}.

In the present work we have chosen the La$_{2-x}$Sr$_{x}$CuO$_{4}$ (LSCO)
solid solution ($x=0.03\div 0.35$). This compound exhibits full range of
behaviors versus chemical composition, which is characteristic of the
layered copper-oxide superconductors. The carrier concentration may be
controlled by the Sr concentration $x$. One could therefore investigate the
doping dependence from the semiconducting region ($x\lesssim 0.05$), through
the underdoped ($0.05\lesssim x\lesssim 0.17$) and overdoped ($0.17\lesssim
x\lesssim 0.30$) superconducting regions, up to the heavily doped metallic
region with no superconductivity ($x\gtrsim 0.30$) \cite{Devaux}. Moreover,
LSCO has a simple crystal structure with single CuO$_{2}$ layers. It has
neither Cu-O chains as in YBa$_{2}$Cu$_{3}$O$_{7-\delta }$ nor complicated
modulation of the separating and spacing layers as in Bi- and Tl-based
materials. All our samples have been characterized by X-ray and electrical
resistivity measurements.

\section{Experimental}

\subsection{Samples}

Polycrystalline samples of La$_{2-x}$Sr$_{x}$CuO$_{4}$ were produced
following the standard solid state technique from high purity La$_{2}$O$_{3}$%
, SrCO$_{3}$ and CuO substrates. The powders were mixed and prefired in air
at 950 $^{\circ }$C for 24 h. After pulverization, the materials were mixed,
pressed and sintered at 1000 $^{\circ }$C for 60 h. Then they were
regrounded, pelletized and, except one sample, refired at 1050 $^{\circ }$C
for 72 h in oxygen under pressure of 1 bar. Only the sample of La$_{1.65}$Sr$%
_{0.35}$CuO$_{4}$ was sintered in oxygen under the pressure of 300 bar at
1000 $^{\circ }$C for 48 h. All products were confirmed to be single phase
by powder X-ray diffraction. The values of the critical temperature measured
by the electrical resistivity measurements are shown in Table 1.

Table 1. Critical temperatures of the La$_{2-x}$Sr$_{x}$CuO$_{4}$ samples.

\begin{tabular}{lllllllll}
$x$ (Sr) & 0.03 & 0.05 & 0.10 & 0.15 & 0.20 & 0.25 & 0 30 & 0.35 \\ 
$T_{c}$ [K] & 0 & 0 & 28.6 & 36.0 & 30.2 & 18.0 & 10.2 & 0
\end{tabular}

\subsection{Choice of the experimental configuration}

The Ettingshausen effect is usually very small, hence it might be easily
overridden by other thermal effects, as the Joule and Thomson effects.
Especially, the Joule heat may particularly hinder the detection of the
Ettingshausen effect. When measuring this effect a strong electrical current
should be passed through the sample thus evolving a large amount of energy
at the electrical contacts, which electrical resistance is usually much
higher than that of the sample. Therefore, some places of the sample should
be thermally anchored to a large mass of constant temperature to carry the
Joule heat away. On the other hand, the sample should be located in
adiabatic conditions since the thermal gradient is the quantity to be
measured. For the above reason we started from computer modelling of
different possible experimental setups, with different patterns of sample
anchoring (e.g. on corners, along the longest side, etc.). In all cases we
have supposed that the sample should have a shape of a flat slab with the
shortest dimension parallel to the magnetic field, since the temperature
difference due to the Ettingshausen effect is inversely proportional to that
dimension (in analogy to the Hall effect). The configuration we have finally
chosen is presented Fig. 2.

The ends of the sample (A) were attached to two copper bars (B) using the
Au:In alloy and silver paint, making both good electrical and thermal
contacts. The typical contact resistance amounted to 0.5$\div $2 $\Omega $
and was much higher than the sample resistance. The bars (B) were
electrically insulated from the copper support (C), but their good thermal
connection to the support was ensured by a relatively large area of the
contacts. The sample was also surrounded by a thick copper screen (D)
screwed on the support (C). The carbon-glass thermometer (E) was \ located
within support (C). The whole assembly was placed in a gas-flow cryostat. A
differential copper-constantan-copper thermocouple was attached to the sides
of the sample. This type of thermocouple has relatively large sensitivity at
room temperature ($\alpha =40.5$ $\mu $V/K). Moreover, since only the middle
segment was made of constantan, the total resistivity of the thermocouple
amounted to few ohms only, reasonable reducing the thermal voltage noise.
Next advantage was that only copper leads were connecting the thermocouple
junctions on the sample with the Keithley 182 nanovoltmeter input, thus
avoiding all detrimental EMF's (the only lead solderings were thermally
anchored to the support C). No detectable influence of the magnetic field on
the thermocouple voltage have been found.

Providing that the length of the sample is much greater than its width, in
our configuration the Ettingshausen coefficient $P$ may be calculated
directly from the definition:

\[
\Delta T_{Ett}=PJ_{x}B_{z}/d 
\]
where $\Delta T_{Ett}$ is the temperature difference between both sides of
the sample due to the Ettingshausen effect, $J_{x}$ is the electrical
current, $B_{z}$ is the magnetic induction and $d$ is the sample thickness.
All our samples had thickness of 0.25-0.35 mm, width of 2.5-3 mm and length
of 9-10 mm.

\subsection{Influence of the Joule effect}

The temperature distribution on the sample surface due the Ettingshausen
effect was presented in Fig. 3a. Our calculations have shown that in the
centre of the sample this distribution is independent on the temperature
distribution caused by the Joule effect, which is shown in Fig. 3b. In other
words, the difference $\Delta T_{Ett}$ remains unaffected by the Joule heat,
even if the values of the temperature gradient along the sample are much
higher then the transversal gradient. The reason is that the gradients
produced by the two effects are perpendicular. However, an inevitable
mismatching of the thermocouple junctions positions causes that the Joule
effect also may significantly contribute to the total temperature difference
measured:

\[
\Delta T=T_{2}-T_{1}=\pm \Delta T_{Ett}+\Delta T_{Joule} 
\]
Depending on particular mounting, the $\Delta T_{Joule}$ difference may be
of any sign. Its value, which is approximately proportional to $J^{2}$, may
be much higher then $\Delta T_{Ett}$. Fortunately, the $\Delta T_{Ett}$
difference changes its sign upon reversing the direction of both electrical
current and magnetic field, whereas $\Delta T_{Joule}$ difference does not.
This feature was used to extract the Ettingshausen effect from the
background.

\subsection{Influence of the Thomson effect}

There is another effect interfering with determination of the Ettingshausen
temperature difference $\Delta T_{Ett}$ - namely the Thomson effect. Because
of our experimental arrangement a temperature gradient along the sample is
present (see Fig. 3a), so the total heat $q$ produced per time unit in the
sample (without magnetic field) consists of two components:

\[
\dot{q}(x)=J_{x}^{2}\rho -\mu _{T}J_{x}\frac{dT(x)}{dx} 
\]
The first term is the Joule heat, the second - the Thomson heat; $\rho $
denotes the electrical resistivity and $\mu _{T}$ - the Thomson coefficient.
According to our knowledge, the values of the Thomson coefficient for HTC
materials are unknown. However, they should be of the order of the Seebeck
coefficient, because of the Kelvin relation, $\mu _{T}=TdS/dT$. The Thomson
effect causes that the temperature distribution along the sample slightly
changes upon changing the direction of the electrical current - see Fig. 4.
Thus, the total temperature difference measured crosswise the sample should
be expressed as:

\[
\Delta T=T_{2}-T_{1}=\pm \Delta T_{Ett}+\Delta T_{Joule}\pm \Delta
T_{Thomson} 
\]
The $\Delta T_{Thomson}$ difference changes sign upon changing the direction
of the electrical current, but, fortunately, it is insensitive to the
magnetic field. Hence, the measurements of $\Delta T$ in function of
magnetic field still give a chance to extract the effect of interest, $%
\Delta T_{Ett}$.

In the course of our experiments we have sometimes observed that during
prolonged measurements on the same sample the values of the $\Delta
T_{Joule} $ and $\Delta T_{Thomson}$ changed significantly (including the
sign change for $\Delta T_{Thomson}$), whereas the\ $\Delta T_{Ett}$
remained unaffected. This observation is not surprising since $\Delta
T_{Joule}$ and $\Delta T_{Thomson}$ are just a result of the mismatching of
the thermocouple junctions which can be effectively changed by thermocycling.

\subsection{Influence of the electrical contacts heating and of the gas
cooling}

The resistance of the sample contacts was usually approximately one order of
magnitude higher than that of sample itself, so most of the electrical
energy was released within the contacts. The only consequence for the
temperature distribution within the sample is that the curves shown in Fig.4
are shifted upwards, with no change of both the longitudinal and transversal
temperature gradients. However, due to relatively large electrical current
values we were using (up to 0.5 A) the sample would be heated up by several
kelvins in respect of its surrounding (i.e. the support C with thermometer
and the screen D, see Fig. 2). Therefore, in order to keep the temperature
of the sample possibly close to that of thermometer we decided to perform
all measurements in helium gas atmosphere under ambient pressure, despite
that it will influence all temperature gradients in the sample and its
support. We can assume that the gas has the temperature of the support.
Thus, the gas is colder than the sample heated by the electrical contacts.
Therefore, one could expect that the longitudinal temperature distributions
presented in Figures 3a and 4 would be substantially flatten, or even
inverted, depending on the ratio of the Joule heating to the gas cooling
efficiency. The diminishing of the transversal gradient due to the
Ettingshausen effect depends only on the gas cooling efficiency. Therefore,
it is possible to set the experimental conditions so that the gas cooling is
weak enough to do not disturb the Ettingshausen effect significantly, but to
reduce substantially the detrimental longitudinal gradients. Indeed, in
tests, in which one of the thermocouple junctions was moved onto the bar B,
we have observed that for low values of the measurement current the middle
of the sample may be colder than its ends (see Fig. 5). Moreover, reducing
high temperature gradients by gas cooling should also result in diminishing
the influence of non-linear energy exchange by radiation.

To check the influence of the gas cooling on the results we have performed
several measurements changing pressure of helium gas in the range of 30$\div 
$1030 mbar. It has been observed that the value of the measured
Ettingshausen coefficient is insensitive to the pressure within the
measurement error. In contrast, the reduction of the pressure from 1030 mbar
to 30 mbar resulted in increase of the $\Delta T_{Joule}$ difference by
approximately 3 times. The $\Delta T_{Thomson}$ difference was also found to
be quite sensitive to the pressure, even that sign changes were observed.

\subsection{Measurements strategy}

A typical measurements sequence without magnetic field is presented in Fig.
6. For each electrical current value several readings of the thermocouple
voltage were taken, with alternatively changing the current direction. A
delay of 60-120 seconds was applied after each current reversal to allow the
steady-state to build up. The average value of the measured voltage for a
given current value ($\Delta T_{Joule}$) was increasing approximately
proportionally to $J^{2}$, as it might be expected for the Joule effect (see
the upper insert). Changing the current direction resulted in a smaller
voltage variations, which amplitude was found to be dependent on the
electrical current value (see the lower insert). We suppose that the Thomson
effect is responsible for these variations. In this paper we will call this
amplitude $\Delta U_{\pm J}$ (an amplitude of the thermocouple voltage
variations due to reversing of the electrical current). The Thomson
temperature difference may be calculated from the equation:

\[
\Delta U_{\pm J}=\alpha (\Delta T_{Thomson}) 
\]
where $\alpha $ is the thermocouple sensitivity. As expected from our
computer simulations, we have not observed any universal dependence of $%
\Delta U_{\pm J}$ on electrical current. For some cases $\Delta U_{\pm J}$
changed its sign for a certain value of the current (with no magnetic field).

Application of the magnetic field causes the Ettingshausen effect and now
the amplitude $\Delta U_{\pm J}$ is a sum of two contributions:

\[
\Delta U_{\pm J}=\alpha (\Delta T_{Thomson}+\Delta T_{Ett}) 
\]
A typical measurement sequence in magnetic field and with constant
electrical current is shown in Fig. 7. The influence of the Ettingshausen
effect is clearly visible. The measured amplitude $\Delta U_{\pm J}$
strongly depends on the field strength. The insert demonstrates the values
of $(\Delta T_{Thomson}+\Delta T_{Ett})=$ $\Delta U_{\pm J}/\alpha $ for
different values of magnetic field fitted by a linear function. The
Ettingshausen coefficient is calculated from the slope $A$ of this line
using the formula:

\[
P=A\frac{d}{J} 
\]
The linearity of the dependence of the Ettingshausen effect on electrical
current value may be checked by plotting the differences $\Delta U_{\pm
J}(B)-$ $\Delta U_{\pm J}(B=0)$ versus current (see Fig. 8).

Taking into account all the above considerations we have developed a
measurements strategy which is illustrated in the Fig. 9. For a particular
sample we started from measuring of the current dependencies of the $\Delta
U_{\pm J}$ amplitude at $B$=+8, 0 and -8 T. Hence, we were able to check the
linearity of $\Delta U_{\pm J}$ with the electrical current value. Then, we
chose a value of current for measuring the field dependency of $\Delta
U_{\pm J}$ amplitude. If possible, we chose a value of $J$ for which $\Delta
U_{\pm J}(B=0)$ was close to zero, in order to avoid the influence of
another interfering effect connected with the Righi-Leduc phenomenon, which
is defined as:

\[
\nabla T_{RL}=S_{RL}B_{z}\nabla T_{x} 
\]
where $S_{RL}$ is the Righi-Leduc coefficient. This effect may result in
another contribution to the transverse thermal gradient if a longitudinal
thermal gradient is present. Fortunately, the lack of the Thomson effect is
an indication of small longitudinal gradients and in this case the
Righi-Leduc effect may be neglected. Moreover, to reduce additionally the
influence of the Joule and Thomson effects all measurements were performed
at $\sim $1 bar of helium gas.

\section{Results and Discussion}

The method described in this paper allows measurements of Ettingshausen
effect of typical for metals, small value. The effect consists in the
transversal thermal gradient appearing due to the electrical current flow in
the presence of the perpendicular magnetic field. Since this weak thermal
effect may be easily overridden by the Joule and Thomson effects a special
measures have been taken to extract it from the background. The sample ends
were thermally anchored to a large mass to carry away the Joule heat. To
eliminate the influence of the longitudinal thermal gradients due to the
Joule and Thomson effects the odd symmetry of the Ettingshausen temperature
difference in respect with the direction of the magnetic field and
electrical current have been exploited.

The values of the measured Ettingshausen coefficients for La$_{2-x}$Sr$_{x}$%
CuO$_{4}$ are shown in. Fig. 10. For each composition 2 or 3 samples were
investigated. For each sample several measurement runs were carried out and
the results were averaged. All the coefficients are of the order of 10$^{-7}$
m$^{3}$K/J, what is typical value for good metals, in contrast to the
semiconductors, where values higher by several orders of magnitude are
expected (see the Introduction). The Ettingshausen coefficient for La$_{2-x}$%
Sr$_{x}$CuO$_{4}$ changes the sign: it is positive only for low Sr
concentration ($x$ = 0.03 and 0.05), for all other compositions negative
values were observed. This is also in contradistinction to the situation of
semiconductors, where only positive values are predicted by the theory \cite
{Paranjape} and found experimentally (see Introduction). The overall weak
compositional dependence is somehow surprising, since the other transport
coefficients for La$_{2-x}$Sr$_{x}$CuO$_{4}$ were found to be strongly
dependent on Sr/La substitution. It was shown \cite{Suzuki,Takagi} that for
the compositional range $x$ $=0.05\div 0.35$ the Hall coefficient decreases
by more than two orders of magnitude. Similar strong variation with
composition was found for thermoelectric power \cite{Cooper,Goodenough}.

\bigskip

\section{Acknowledgments}

The work was supported by the Polish State Committee for Scientific Research
under contract No. 2PO3B 11613.

\newpage {\LARGE Figure captions}

{\bf Figure1} \ The sign convention for a positive Ettingshausen effect.

{\bf Figure 2} An experimental setup used for measurements of the
Ettingshausen effect. See text for details.

{\bf Figure 3} \ The temperature distribution on the sample area due the
Ettingshausen effect (a) and the temperature distribution along the sample
due to the Joule effect (b). The mismatching of the thermocouple junctions
positions was exaggerated to indicate how the Joule effect may contribute to
the measured temperature difference $\Delta T=T_{2}-T_{1}$. Calculated by
the finite-elements method.

{\bf Figure 4} \ The temperature distribution along the sample due to the
Joule and Thomson effects for two directions of the electrical current. The
difference between two curves was exaggerated for sake of clarity.
Calculated by the finite-elements method.

{\bf Figure 5} \ A realistic temperature distribution along the sample for
different values of the measurement current (calculated by the
finite-elements method). The Joule and Thompson effects within the sample,
the Joule effect in the electrical contacts and the cooling by the
surrounding gas have been taken into account. The sample ends are warmer in
respect with the support due to the Joule heating of the contacts. It is
assumed that the gas has the same temperature as the support. The insert
shows the results of the tests, in which one of the thermocouple junctions
was moved onto the bar B (see Fig. 2) and the longitudinal gradient $\Delta
T_{l/2}$ has been measured.

{\bf Figure 6} \ An exemplary measurement sequence performed on La$_{1.75}$Sr%
$_{0.25}$CuO$_{4}$\ sample at room temperature without magnetic field. The
thermocouple voltages are plotted versus the successive reading numbers. The
direction of the electrical current was alternatively changed from reading
to reading. Our convention is that $J$\ is positive for odd reading numbers,
which are denoted by filled circles. Even numbers are denoted by empty
circles. The upper insert shows the average voltage for each value of the
current versus the current, i.e. the result of the Joule effect (the
thermocouple sensitivity is denoted by $\alpha $). The variations of the
voltage with the current direction are due to the Thomson effect. The
amplitude of these variations $\Delta T_{Thomson}=\Delta U_{\pm J}/2\alpha $
versus current value is shown on the lower insert.

{\bf Figure 7}\ \ An exemplary measurement sequence performed on La$_{1.75}$%
Sr$_{0.25}$CuO$_{4}$\ sample at room temperature for a constant electrical
current and in varied magnetic field (see also the caption of Fig. 6). A
plot of the amplitude ($\Delta T_{Thomson}+\Delta T_{Ett})=\Delta U_{\pm
J}/2\alpha $\ versus $B$\ is shown in the insert.

{\bf Figure 8} \ A plot of the amplitude ($\Delta T_{Thomson}+\Delta
T_{Ett})=\Delta U_{\pm J}/2\alpha $ versus electrical current for different
values of the magnetic field. The insert shows the differences $(\Delta
U_{\pm J}(B)-$\ $\Delta U_{\pm J}(B=0))/2\alpha =\pm \Delta T_{Ett}$ ($%
\alpha $ is the thermocouple sensitivity).

{\bf Figure 9} \ A sketch of the experimental procedure for measurements of
the Ettingshausen effect. At each point on the ($B$,$J$) plane several
readings of the transversal temperature gradient have been taken for both
current directions (see Figures 6 and 7).

{\bf Figure 10} The room temperature Ettingshausen coefficients for La$%
_{2-x} $Sr$_{x}$CuO$_{4}$. The error bars denote the scatter of experimental
values obtained for a particular composition.

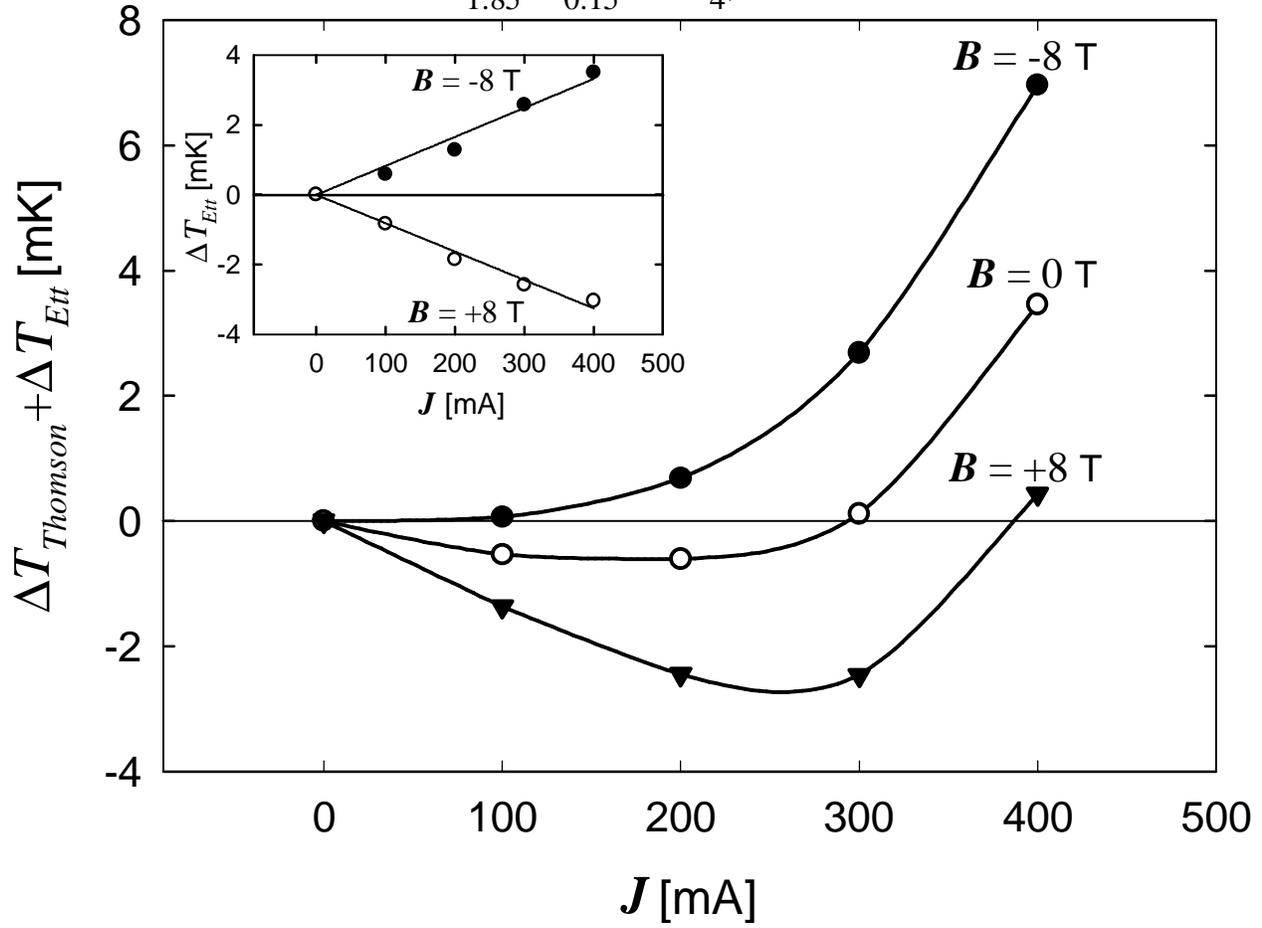
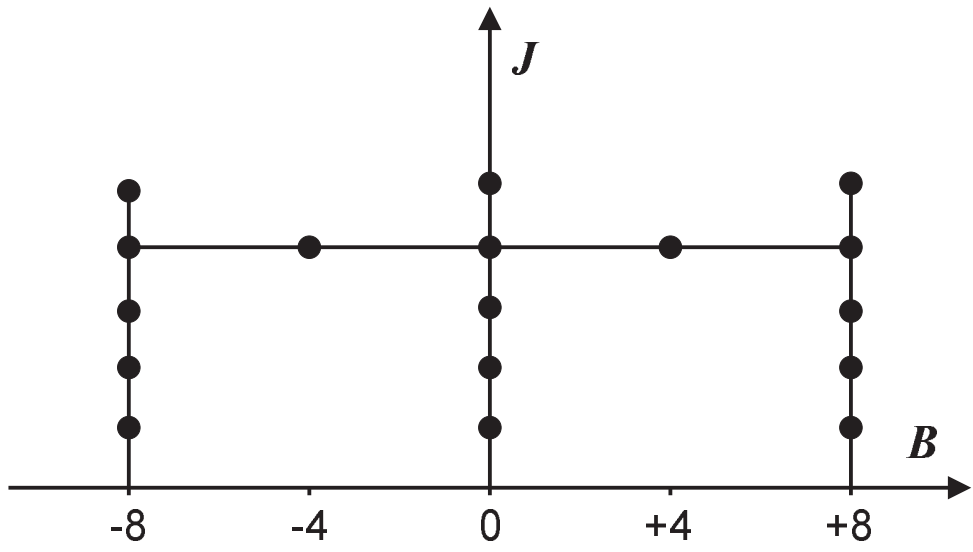
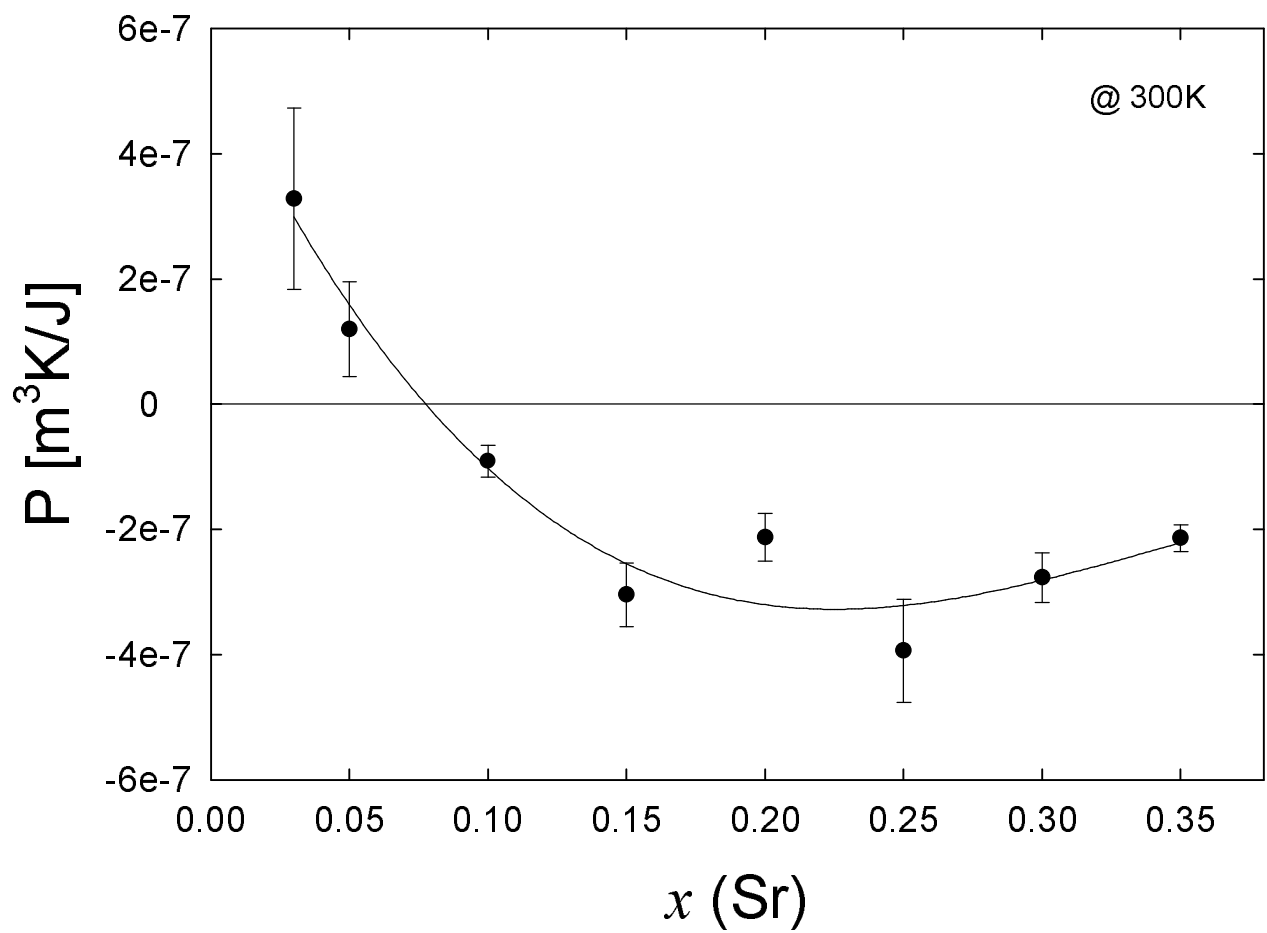
\end{document}